\documentclass[12pt]{iopart}
\usepackage{harvard}
\usepackage{hyperref}
\usepackage{titlesec}
\usepackage{diagbox}
\usepackage{booktabs}
\usepackage{multirow}
\usepackage[english,printdayoff]{isodate}
\usepackage[bf]{caption}
\usepackage{subcaption}
\usepackage[shortlabels]{enumitem}
\titleformat{\paragraph}
{\normalfont\normalsize\bfseries}{\theparagraph}{1em}{}
\titlespacing*{\paragraph}
{0pt}{3.25ex plus 1ex minus .2ex}{1.5ex plus .2ex}

\expandafter\let\csname equation*\endcsname\relax
\expandafter\let\csname endequation*\endcsname\relax

\usepackage{amsmath}
\usepackage{siunitx}
\usepackage{amsthm}
\theoremstyle{definition}

\usepackage{enumitem}
\usepackage{graphicx}
\usepackage{tikz}

\begin{document}
\title[A GPU-based MCO algorithm for HDR brachytherapy]{A GPU-based multi-criteria optimization algorithm for HDR brachytherapy}

\author{C\'edric B\'elanger$^{1,2\ddag}$, Songye Cui$^{1,2\ddag}$, Yunzhi Ma$^{2}$, Philippe Despr\'es$^{1,2}$, J. Adam M. Cunha$^{3}$, Luc Beaulieu$^{1,2}$}
\address{$^1$Department of Physics, Engineering Physics and Optics and Cancer Research Center, Universit\'e Laval, Quebec City, QC, G1V 0A6, Canada}
\address{$^2$Department of Radiation Oncology and Research Center of CHU de Qu\'ebec - Universit\'e Laval, Quebec City, QC, G1R 2J6, Canada}
\address{$^{3}$Radiation Oncology, University of California, San Francisco, CA 94115, USA}
\address{$^{\ddag}$Co-first authorship}
\ead{Luc.Beaulieu@phy.ulaval.ca}

\begin{abstract}
Currently in HDR brachytherapy planning, a manual fine-tuning of an objective function is necessary to obtain case-specific valid plans. This study intends to facilitate this process by proposing a patient-specific inverse planning algorithm for HDR prostate brachytherapy: GPU-based multi-criteria optimization (gMCO).

Two GPU-based optimization engines including simulated annealing (gSA) and a quasi-Newton optimizer (gL-BFGS) were implemented to compute multiple plans in parallel. After evaluating the equivalence and the computation performance of these two optimization engines, one preferred optimization engine was selected for the gMCO algorithm. Five hundred sixty-two previously treated prostate HDR cases were divided into validation set (100) and test set (462). In the validation set, the number of Pareto optimal plans to achieve the best plan quality was determined for the gMCO algorithm. In the test set, gMCO plans were compared with the physician-approved clinical plans.

Our results indicated that the optimization process is equivalent between gL-BFGS and gSA, and that the computational performance of gL-BFGS is up to 67 times faster than gSA. Over 462 cases, the number of clinically valid plans was 428 (92.6\%) for clinical plans and 461 (99.8\%) for gMCO plans. The number of valid plans with target $V_{100}$ coverage greater than 95\% was 288 (62.3\%) for clinical plans and 414 (89.6\%) for gMCO plans. The mean planning time was \SI{9.4}{\second} for the gMCO algorithm to generate \num{1000} Pareto optimal plans.

In conclusion, gL-BFGS is able to compute thousands of SA equivalent treatment plans within a short time frame. Powered by gL-BFGS, an ultra-fast and robust multi-criteria optimization algorithm was implemented for HDR prostate brachytherapy. Plan pools with various trade-offs can be created with this algorithm. A large-scale comparison against physician approved clinical plans showed that treatment plan quality could be improved and planning time could be significantly reduced with the proposed gMCO algorithm.

\end{abstract}

\maketitle 
\section{Introduction}

About 52.3\% of non-skin cancer patients receive radiation therapy during the course of their illness~\cite{CITR17,DEVI15,DELA05}. The most common radiation therapy treatment particle type used is the photon, which can be delivered either externally from a medical linear accelerator (External Beam Radiation Therapy - EBRT) or internally from an inserted small radioactive source (brachytherapy, high dose rate (HDR) or low dose rate (LDR)).

Dose prescriptions in modern radiation treatment planning contain both tumor and healthy organ objectives. These objectives are often conflicting and can be generalized as: treating the tumor with high radiation dose and sparing the healthy organs with low radiation dose. Computerized treatment planning systems were used to formulate clinical prescriptions into a mathematical optimization problem, and to find treatment plans that well presented these prescriptions with treatment facilities.

However, most available algorithms are not inherently patient-specific in a sense that manual re-plannings are usually inevitable to find a clinically acceptable plan for each patient. As a result, the planning procedure can be time consuming and the planning output is planner dependant~\cite{MOOR11,NELM12,WU09a}.

Several patient-specific inverse planning algorithms such as knowledge-based planning (KBP), auto-planning (AP) and multi-criteria optimization (MCO) have been proposed in EBRT. In KBP, one plan is created for a new case by searching in a prior physician-approved plan dataset based on the geometric features~\cite{MOOR11,WU11,WU09a,PETI12}. In AP, a clinical plan can be obtained by interactively and automatically adapting objectives, constraints and dose shaping contours~\cite{HAZE15}. In MCO, a plan pool is constructed by generating plans with various trade-offs on Pareto surfaces~\cite{CRAF06,TEIC11}. Similar studies can also be found in brachytherapy~\cite{MEER18,SHEN18,ZHOU17,CUI18,CUI18a}.

Our prior studies~\cite{CUI18,CUI18a} showed that a patient-specific treatment plan can be created without any user interventions in HDR prostate brachytherapy. However, the optimization engine of these studies was stochastic, and was implemented on CPU hardware~\cite{CUI18,CUI18a}. As a result, the algorithm inevitably involved an intensive computation (\SI{41}{\second}), which may restrain its application in clinical practice, because the patient is under general anesthesia in the operating room waiting for the treatment to be delivered. 

The capability of graphics processing unit (GPU) architecture in reducing calculation time in medical physics were reviewed in~\cite{PRAT11a,JIA14,DESP17}. The purpose of this study is to propose an ultra-fast patient-specific inverse planning algorithm on GPU for HDR brachytherapy.

\section{Methods and Materials}

This section begins with a detailed description of experimental setups including patient selection, mathematical formulations and computational specifications. Next, two inverse planning optimization engines were implemented on GPU architecture to calculate multiple plans in parallel and to populate the Pareto surfaces. Powered by the preferred optimization engine, a GPU-based multi-criteria optimization algorithm (gMCO) which is able to automatically generate clinical plans was proposed to eliminate the re-planning problem in HDR brachytherapy. In the end, a comprehensive comparison, including dosimetric performance as well as planning time, between clinical plans and gMCO plans was made.

\subsection{Experimental setup}

\subsubsection{Patient selection}

An anonymous dataset that contains 562 prostate cancer patients who received an HDR brachytherapy treatment as a boost to EBRT from \daterange{2011-04-13}{2016-07-06} at our institution was studied. This dataset incorporates the cases studied in prior works~\cite{EDIM19,CUI18,CUI18a}. Among the dataset, 100 random cases (validation set) were used to determine the number of Pareto optimal plans with the gMCO algorithm, and 462 random cases (test set) were used in the performance evaluation of the gMCO generated plans.

After inserting 16-18 plastic catheters into the prostate under a transrectal ultrasound guidance, the anatomy of these patients was obtained from CT scans. Organ structures (prostate, urethra, bladder and rectum) were delineated and were imported into a commercial treatment planning system (Elekta Oncentra Brachy IPSA, Veenendaal, The Netherlands). The prescription was to deliver 15~Gy in a single fraction to the prostate. Plans were delivered using a Flexitron afterloader (Elekta Brachy, Veenendaal, The Netherlands) with an Ir-192 radioactive source.

The dwell positions were extracted from the DICOM-RT files of clinical plans, and the mean number of active dwell positions ($N_\text{act}$) used for the optimization was 171 (range:102-385). The mean number of dose calculation points ($N_\text{pnt}$) used for the optimization was \num{5913} (range:\num{2753}-\num{15998}), and the mean number of dose calculation points used for the dose-volume histogram (DVH) computations was \num{31039} (range:\num{11451}-\num{66089}).

\subsubsection{Quadratic objective function formulation}
\label{sec:math}

Inverse Planning Simulated Annealing (IPSA)~\cite{LESS01} was used as a dose optimization engine in our prior studies~\cite{CUI18,CUI18a}. In IPSA, piecewise linear objective functions were solved with simulated annealing~\cite{LESS01}, a stochastic optimizer. These objective functions were constructed with a population based planning template called a class solution~\cite{CUI18,CUI18a}. 

In order to implement an efficient optimizer, one option is to replace the stochastic optimizer with a gradient-based optimizer. Therefore, it may be necessary to replace the IPSA linear piecewise objective functions with piecewise quadratic objective functions, so that the first derivative (gradient) of the objective function is continuous. Quadratic objective functions are usually solved with gradient-based optimizers in radiation therapy~\cite{MILI02,LAHA03a,LAHA03c,MEN09}. 

The dose at the $i^{th}$ dose calculation point in the $j^{th}$ organ, denoted by $d_{ij}$, is described in equation~(\ref{eq:dose})
\begin{equation}
    d_{ij} = \sum_{l=1}^{N_\text{act}} \dot{d}_{ijl}t_l
    \label{eq:dose}
\end{equation}
where $\dot{d}_{ijl}$ is the dose rate contribution of the $l^{th}$ dwell position to the $i^{th}$ dose calculation point in the $j^{th}$ organ, and $t_l$ is the dwell time of the $l^{th}$ dwell position. In order to avoid negative dwell times, new decision variables called dwell weight ($x_{l} = t_l^{1/2}$) were introduced as in~\cite{MILI02,LAHA03a}. With this substitution, the dwell times are always non-negative ($t_l = x_l^{2}$). 

The piecewise quadratic objective function $f_{ij}$ at the $i^{th}$ dose calculation point of the $j^{th}$ organ is given in equation~(\ref{eq1})

\begin{equation} \label{eq1}
f_{ij}(d_{ij}) =
\begin{cases}
	w_{\text{min}}\cdot(D_{\text{min}}-d_{ij})^{2} & d_{ij} < D_{\text{min}} \\
	0 & D_{\text{min}} \leq  d_{ij}  \leq  D_{\text{max}} \\
    w_{\text{max}}\cdot(d_{ij}-D_{\text{max}})^{2}  & d_{ij} > D_{\text{max}}\,.
\end{cases}
\end{equation}
Variables $D_{\text{min}}$ and $D_{\text{max}}$ are the underdose limit and the overdose limit respectively, and variables $w_{\text{min}}$ and $w_{\text{max}}$ are the corresponding weights. The corresponding gradient function $g_{ij}$ of equation~(\ref{eq1}) is described in equation~(\ref{eq2})

\begin{equation} \label{eq2}
g_{ij}(x_l) = \frac{\partial{f_{ij}}}{\partial{x_{l}}} =
\begin{cases}
	4\cdot \dot{d}_{ijl}\cdot x_l\cdot w_{\text{min}}\cdot(d_{ij}-D_{\text{min}}) & d_{ij} < D_{\text{min}} \\
	0 & D_{\text{min}} \leq  d_{ij}  \leq D_{\text{max}} \\
    4\cdot \dot{d}_{ijl}\cdot x_l\cdot w_{\text{max}}\cdot(d_{ij}-D_{\text{max}})  & d_{ij} >D_{\text{max}}\,.
 \end{cases}
\end{equation}

The single joint MCO objective function to be minimized is defined as a weighted sum in equation~(\ref{eq:obj})

\begin{equation} \label{eq:obj}
F = \sum_{j=1}^{N_\text{O}}w_j\cdot \frac{1}{N_{\text{pnt}, j}}\sum_{i=1}^{N_{\text{pnt}, j}}f_{ij}(d_{ij})
\end{equation}
where $N_\text{O}$ is the number of organs, $N_{\text{pnt}, j}$ is the number of dose calculation points in the $j^{th}$ organ. $w_j$ is a hidden weight applied to the objectives (surface and volume) of the $j^{th}$ organ to introduce trade-off in the solution space around the population-based starting point as in~\cite{CUI18,CUI18a}. The hidden weights are always non negative and their sum is one (because of the weighted sum method).

The original class solution designed for the piecewise linear objective functions~\cite{CUI18,CUI18a} will no longer be appropriate to construct the new quadratic objective functions, and so a new one must be designed (table~\ref{table1}). 

\begin{table}[htbp]
 \centering
 \caption{The class solution to formulate quadratic objective functions (equation~(\ref{eq1})) for \SI{15}{\gray} prostate boost HDR treatment (Surface: surface dose calculation points, Volume: volume dose calculation points).}\label{table1}
  \begin{tabular}{cccccccccc} \hline 
      \multicolumn{1}{c}{\multirow{2}{*}{Organ}} & \multicolumn{4}{c}{Surface} & &\multicolumn{4}{c}{Volume}\\ \cline{2-5}\cline{7-10}
      & $w_\text{min}$ & $D_\text{min}$(Gy) & $D_\text{max}$(Gy) & $w_\text{max}$ 
      && $w_\text{min}$ & $D_\text{min}$(Gy) & $D_\text{max}$(Gy) & $w_\text{max}$ \\ \hline
      Target & 200 & 15 & 22.5 & 80 
      	&& 200 & 15 & 22.5 & 1 \\
	  Urethra & 30 & 14 & 16 & 160 
      && 30 & 14 & 16 & 160\\
      Bladder & 0 & 0 & 7.5 & 60 
      	&& 0 & 0 & 7.5 & 60\\
      Rectum & 0 & 0 & 7.5 & 15& & 0 & 0 & 7.5 & 15\\ \hline
  \end{tabular}
\end{table}

\subsubsection{Computational specifications}

The CPU algorithm was written in \textit{C++}, compiled with \textit{g++} (7.3.0) and executed on a six-core Intel® Xeon® CPU (E5-2620 v3 @ 2.40 GHz). The GPU algorithms were written in CUDA C, compiled with \textit{nvcc} (CUDA toolkit 10.0.130) and executed on an NVIDIA Titan X (Pascal) GPU.

\subsection{GPU-based efficient optimization engines}
\label{GPU_opt}
Previous studies showed that it is feasible to find clinically acceptable treatment plans after exploring Pareto surfaces with MCO approaches~\cite{CRAF06,CUI18}. However, constructing Pareto surfaces could be inefficient, if performed sequentially.

\subsubsection{IPSA on GPU}
\label{GPU}

A traditional CPU-based inverse planning algorithm such as IPSA (or cSA)~\cite{LESS01} can be divided into several serial computing steps (figure~\ref{fig:CPUGPU}). In each step, the same operation is repeated over a large dataset. For example, the following five steps are essential in cSA: 

\begin{figure}[htbp]
\centering
\includegraphics[width=0.9\textwidth]{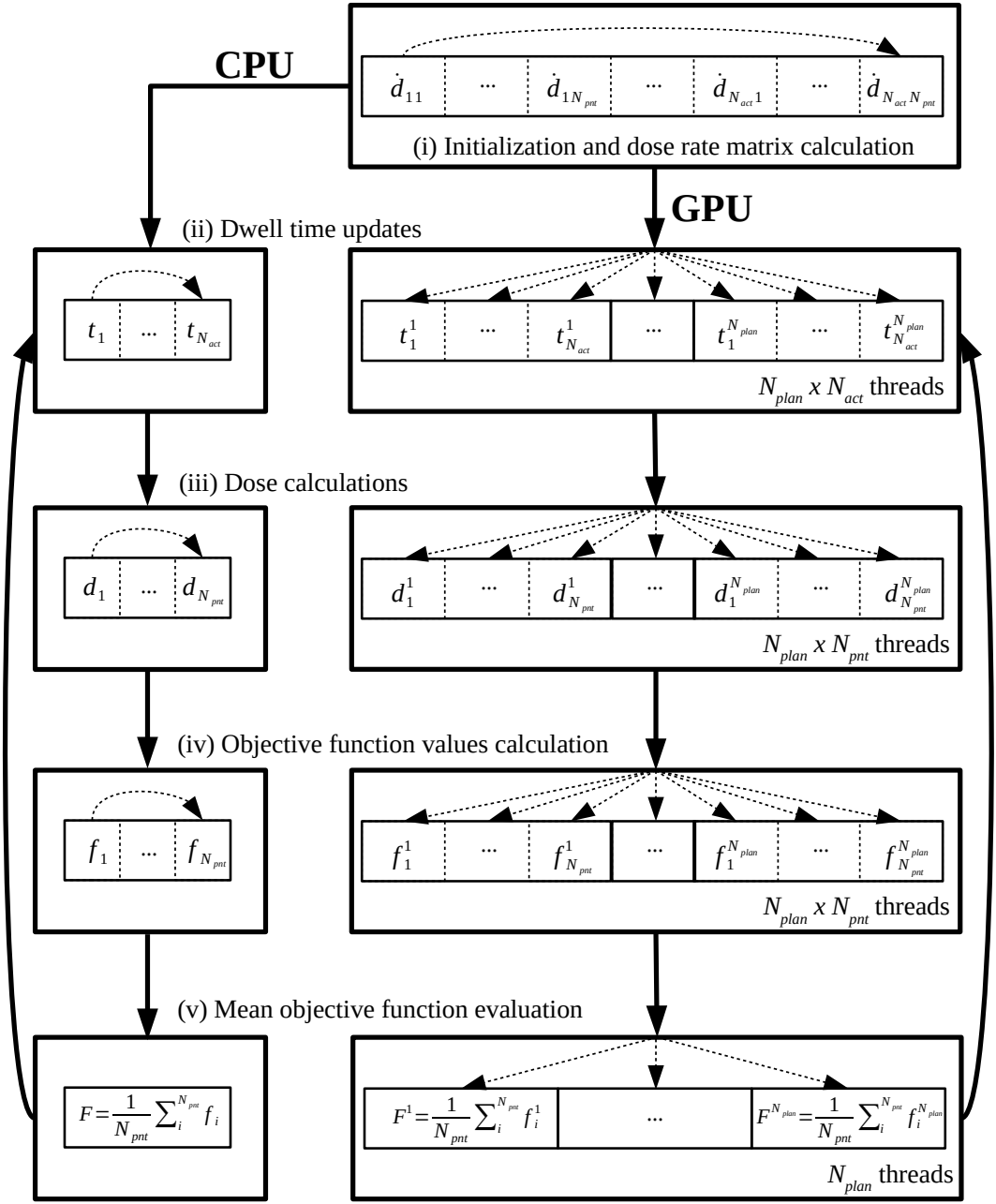}
\caption{Illustration of the iterative procedure to optimize one treatment plan on CPU and $N_\text{plan}$ plans on GPU. In each CPU or GPU iteration, the steps (ii)-(v) are executed sequentially. In each step on the CPU, the operations are executed sequentially in a loop. In each step on the GPU, the operations are executed in parallel on different threads for $N_\text{plan}$ plans. (The superscript
indicates the plan number on GPU).}
\label{fig:CPUGPU}
\end{figure}

\begin{enumerate}[(i)]
 \item Initialization and dose rate matrix calculation (repeated for: $N_\text{pnt}$ dose calculation points $\times$ $N_\text{act}$ dwell positions),
 \item dwell time updates (repeated for: $N_\text{act}$ dwell positions),
 \item dose calculations based on equation~(\ref{eq:dose}) (repeated for: $N_\text{pnt}$ dose calculation points),
 \item objective function values calculation based on equation~(\ref{eq1}) (repeated for: $N_\text{pnt}$ dose calculation points),
 \item mean objective function evaluation based on equation~(\ref{eq:obj}) (repeated for: one accumulation over $N_\text{pnt}$ dose calculation points).
\end{enumerate}

To obtain an optimal solution or a treatment plan, steps (ii)-(v) are iteratively repeated in cSA. Furthermore, in order to explore Pareto surfaces by computing $N_\text{plan}$ treatment plans, it is usually necessary to repeat the aforementioned steps $N_\text{plan}$ times.

To increase the efficiency of MCO approaches, GPU-based IPSA (or gSA) was implemented on GPU architecture to compute treatment plans with various trade-offs in parallel. Two strategies were applied to achieve this purpose.

First, the serial operations computed in each step in cSA were adapted to run in parallel on GPU, so the operations within each step can be executed simultaneously on different threads (figure~\ref{fig:CPUGPU}). Note that in each step on GPU, the computational burden is $N_\text{plan}$ times larger than in the CPU implementation ($N_\text{plan}$ plans on GPU vs. one plan on CPU in figure~\ref{fig:CPUGPU}). However, a performance gain can be achieved with the GPU implementation, as the huge burden of updating the values for all plans in each step is processed in parallel on different threads. To obtain $N_\text{plan}$ optimal solutions or $N_\text{plan}$ treatment plan with the proposed implementation, it is necessary to iteratively repeat steps (ii)-(v) in gSA.



Second, as frequent data transfers between CPU and GPU will slow down the computation, data transfer only occurs twice in gSA: once when preparing the data used for the optimization (CPU to GPU), once more when saving the dosimetric results onto the disk after the optimization (GPU to CPU).

\subsubsection{Deterministic optimizer}

In section~\ref{GPU}, a stochastic optimizer was implemented on CPU and on GPU. To further improve the computational performance, a deterministic optimizer (Limited-memory Broyden-Fletcher-Goldfarb-Shanno, L-BFGS)~\cite{LIU89,BFGS19,JENS13} was introduced to replace the stochastic optimizer. There are two reasons to choose this quasi-Newton optimizer, (1) BFGS and its variants are widely studied in brachytherapy~\cite{MILI02,LAHA03a}, and (2) L-BFGS is widely used in clinic after being integrated in Hybrid Inverse Planning Optimization (HIPO) (Elekta Brachy, Veenendaal, The Netherlands)~\cite{KARA05}.

So far, four optimization engines were implemented: cSA, gSA (simulated annealing on CPU and on GPU), cL-BFGS and gL-BFGS (L-BFGS on CPU and on GPU). The description of L-BFGS implementation on CPU and on GPU is omitted in this study, due to the similarity with the context and figure~\ref{fig:CPUGPU} in section~\ref{GPU}.

\subsubsection{Equivalence between the four optimizers}
\label{sec:convergence}
The equivalence between the four optimization engines was evaluated based on the same objective function (class solution in table~\ref{table1}) as tested over the validation set. For cSA, gSA, and cL-BFGS, one plan using uniform \SI{5}{\second} initial dwell times as a starting point was generated. For gL-BFGS, \num{1000} degenerated plans were calculated to evaluate the convergence of different starting points (randomly distributed between 0 and \SI{10}{\second}). The stopping criteria for cSA and gSA was specified by the number of iterations. The stopping criteria for cL-BFGS and gL-BFGS was specified by the parameter $\epsilon$ (based on the relative variation of the objective function~\cite{MEN09}). To measure the equivalence between the four optimizers, \num{1000000} iterations and $\epsilon = 10^{-7}$ were used as the stopping criteria, because no significant improvements in the objective function were observed.

\subsubsection{Pareto surfaces characterization with gSA and gL-BFGS}
\label{sec:pareto}
Planning efficiency is a key factor when designing an inverse planning algorithm. For SA, a clinically useful stopping criteria (\num{50000} iterations) can be used to reach Pareto surfaces~\cite{CUI18}. For gradient-based method, it is also desirable to find a stopping criteria that can well approximate the Pareto surfaces.

By computing solutions in parallel with various combinations of hidden weights, Pareto surfaces can be populated either with gSA and with gL-BFGS. Such solutions were Pareto optimal, or non-dominated, if no solution that improves any individual objective value without worsening at least one of the other individual objective values exists. A clinically useful stopping criteria was determined for gL-BFGS to approximate the Pareto surfaces, after examining the effect of different stopping criteria (ranging from $\epsilon = 10^{-7}$ to $\epsilon = 10^{-2}$) based on the fraction of non-dominated solutions and the speedup factor of the optimization time for all 100 validation cases.

\subsubsection{Computational performance under clinically useful scenarios}

The benefits of the proposed GPU implementation over a traditional CPU implementation of inverse planning algorithms were explored. Based on the clinically useful stopping criteria, the computational performance of cSA, gSA, cL-BFGS and gL-BFGS were measured against the number of generated plans.

\subsection{Patient-specific multi-criteria optimization algorithm}

Usually, plans obtained with a population-based planning template are not always directly acceptable, and manual weights adjustments are required to obtain a patient-specific deliverable plan. After reviewing the definition of acceptable plans, a GPU-based multi-criteria optimization algorithm (gMCO) powered with gL-BFGS was proposed to eliminate the procedure of manual weights adjustments.

\subsubsection{Plan evaluation}
\label{sec:RTOG}

The schedules of dose fractionation and the evaluation criteria of HDR prostate 
brachytherapy plans may vary between centers~\cite{YAMA12}. According to the Radiation Therapy Oncology Group (RTOG) 0924 protocol~\cite{RTOG16}, RTOG acceptable plans (or valid solutions) can be summarized as follows:
    
\begin{itemize} 
	\item Prostate/Target coverage constraint: $V_{100}$ $\geq$ 90\% of the volume.
    \item Urethra constraint: $D_{10}$ $<$ 118\% of the prescription dose.
     \item Bladder constraint: $V_{75}$ $<$ 1~cc.
    \item Rectum constraint: $V_{75}$ $<$ 1~cc.
\end{itemize}

Note: 

(1)	$V_\text{x}$ refers to the absolute volume that receives x\% of the prescription dose, and $D_\text{x}$ refers to the percent of the prescription dose that covers x\% of the volume.

(2)	In this study, a more stringent set of criteria was introduced. It is designated by the RTOG+ symbol and is the same as the RTOG criteria set except that it specifies a higher target coverage requirement of 95\% for the $V_{100}$. This is usually attainable in the clinic without sacrificing the OAR protection.

\subsubsection{gMCO algorithm}
\label{sec:gMCO}
Compared with our previous studies~\cite{CUI18,CUI18a}, there are three main differences in gMCO: (1) the trade-off between target and urethra is now explored, (2) the Pareto surfaces are widely explored with a large number of plans, as no prior knowledge of the RTOG+ valid solution space is involved, and (3) the validation cases were used to determine the number of parallel plans (from 1 to \num{10000}) needed to achieve high RTOG and RTOG+ acceptance rates with random hidden weights. In gMCO, the parallel plan computations were executed with gL-BFGS. 

\subsection{Comparison between clinical plans and gMCO plans}

A plan pool was created with the gMCO algorithm. One plan was selected from the plan pool and was referred to as the gMCO plan.

The criteria used for plan selection are, in descending order of priority: RTOG+ valid plan, RTOG valid plan, RTOG invalid plan (violates at least one criteria). If multiple RTOG or RTOG+ valid plans existed, the one with a highest target $V_{100}$ was selected. If multiple RTOG invalid plans existed, the one with the lowest bladder and rectum $V_{75}$ (while not violating the criteria for target and urethra) was selected.

\subsubsection{Dosimetric performance}
 
The dosimetric results of clinical plans were retrieved from Oncentra Brachy (Elekta Brachy, Veenendaal, The Netherlands). Dosimetric comparisons between clinical plans and gMCO plans were analyzed for 462 test cases. The overall result was examined based on RTOG and RTOG+ acceptance rates (the criteria of all organs were met). The acceptance rate (i.e. target $V_{100}$, urethra $D_{10}$, bladder $V_{75}$, and rectum $V_{75}$) for each organ was also reported.

\subsubsection{Planning time}

The planning time consists of the time taken for dose calculation points creation, dose rate matrix calculation, optimization, and DVH calculation on GPU. The calculation time of each portion was recorded for gMCO plans. The total planning time was compared between clinical plans and gMCO plans. 

\section{Results}

\subsection{GPU-based optimization engines}

\subsubsection{Equivalence between the four optimizers}

\begin{figure}[htbp]
\centering
\includegraphics[width=0.6\textwidth]{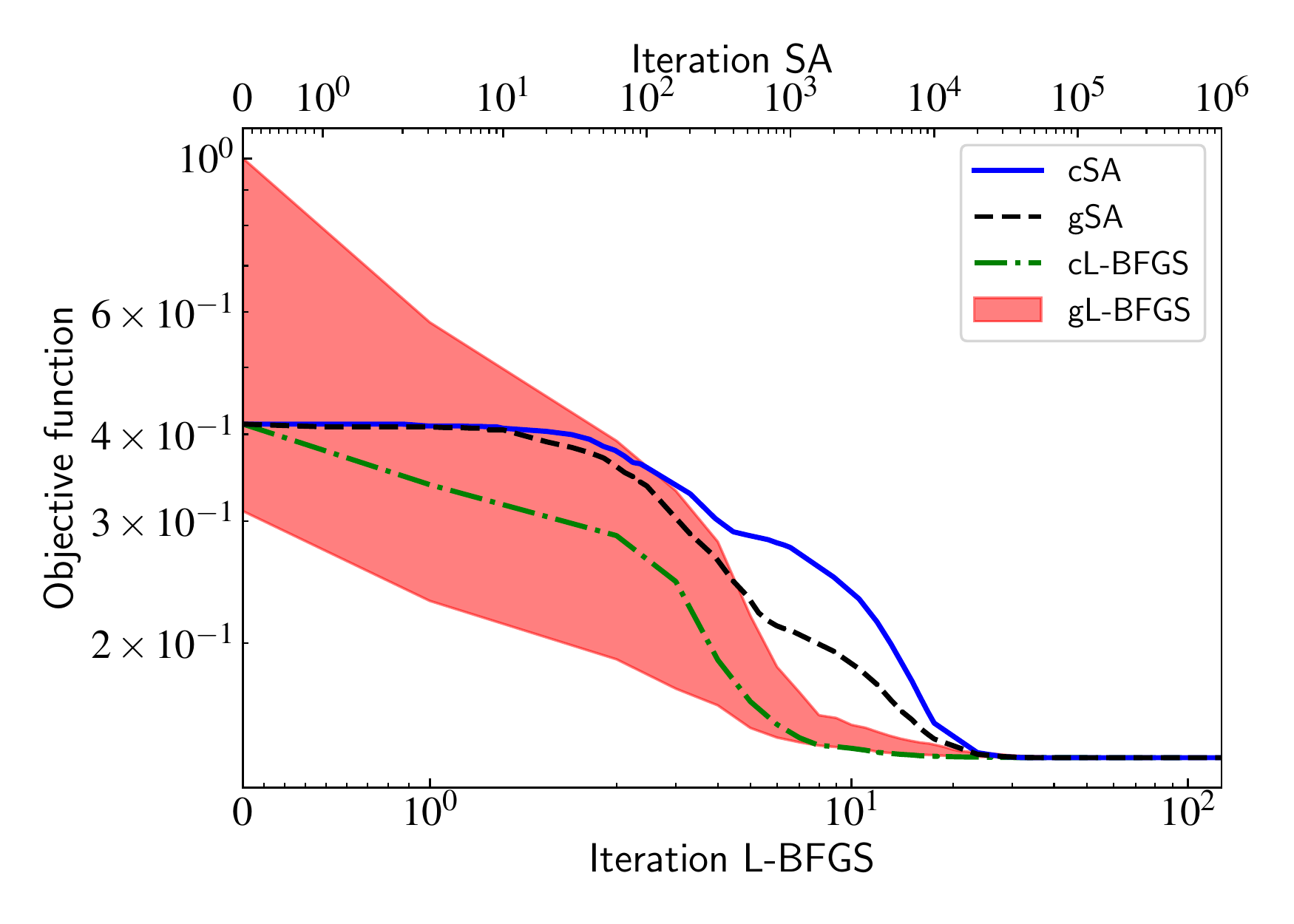}
\caption{Illustration of cSA, gSA, cL-BFGS and gL-BFGS objective function values against the number of iterations for one random validation case. (The difference between CPU and GPU random number generators accounts for the different trajectories for cSA and gSA).}
\label{fig:obj}
\end{figure}

The optimization processes of the four optimizers for one random validation case are illustrated in figure~\ref{fig:obj}. From this figure, (1) gL-BFGS plans obtained with different initial dwell times converge to the SA objective function value, (2) no significant differences (within 0.02\%) in objective function values resulted from the four optimizers were observed. Over all 100 validation cases, similar results were observed, because the final objective function values of the four algorithms were in agreement within 0.2\%.

\subsubsection{Pareto surfaces characterization with gSA and gL-BFGS}

To characterize Pareto surfaces, \num{100000} different solutions were generated with gSA and gL-BFGS (\num{1000} solutions/case for all 100 validation cases). For gSA, the mean fraction of non-dominated solutions was 99.6\% under \num{50000} iterations.

\begin{figure}[htbp]
\begin{subfigure}[b]{0.5\textwidth}
\centering
\includegraphics[width=\textwidth]{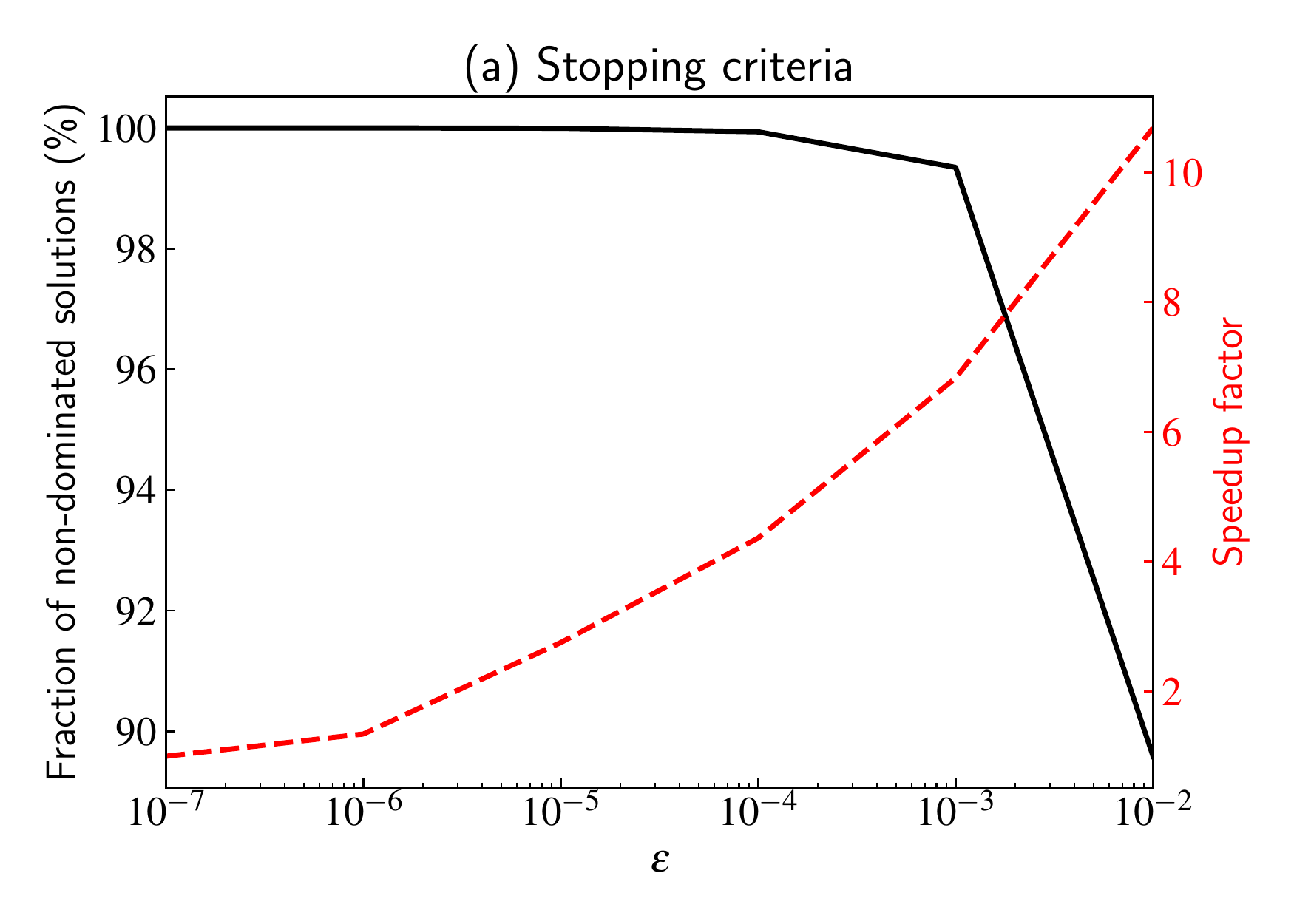}
\end{subfigure}
\begin{subfigure}[b]{0.5\textwidth}
\centering
\includegraphics[width=\textwidth]{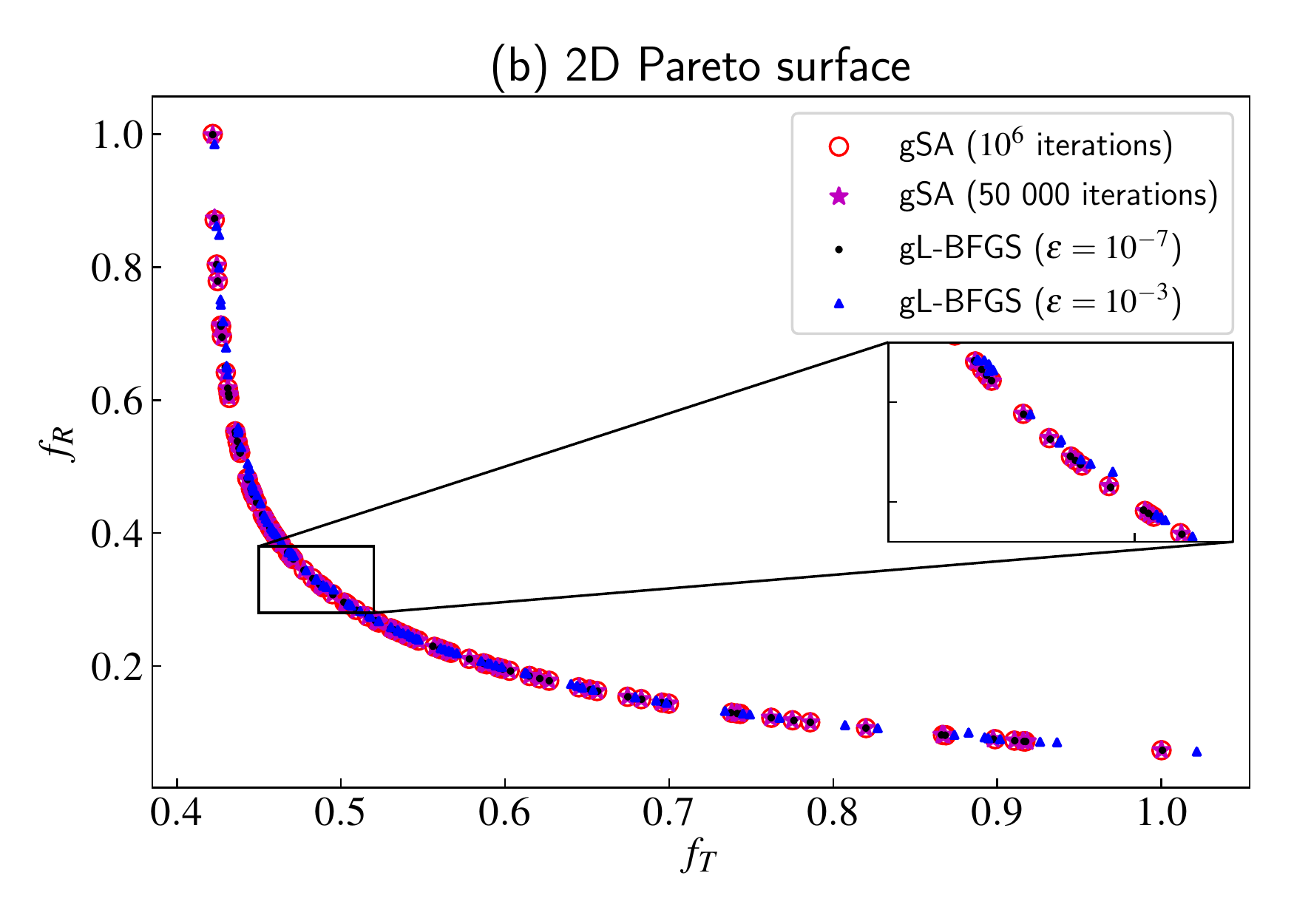}
\end{subfigure}
\caption{(a) Effect of the stopping criteria on the fraction of non-dominated solutions in the Pareto front characterized with gL-BFGS (black solid line) and the speedup factor of the optimization time (red dashed line). The speedup factor are normalized to the values obtained with a stopping criteria of $\epsilon = 10^{-7}$. (b) A comparison of 2D Pareto surface approximations with gSA and gL-BFGS optimization engines for a random case. ($f_T$ is denoted for target individual objective function and $f_R$ is denoted for rectum individual objective function).}
\label{fig:speed_up_factor}
\end{figure}

For gL-BFGS, the results in figure \ref{fig:speed_up_factor}a indicate that the fraction of non-dominated solutions decreased (from 100\% to 89.3\%) as the stopping criteria increased (from $\epsilon = 10^{-7}$ to $\epsilon = 10^{-2}$). On the other hand, the speedup factor in the optimization time increased (from 1 to 10) as the stopping criteria increased (from $\epsilon = 10^{-7}$ to $\epsilon = 10^{-2}$). It should be noted that over 99.3\% of the solutions obtained with a larger stopping criteria ($\epsilon = 10^{-3}$) are Pareto optimal solutions. Given that reaching optimality and a reasonable calculation time are important criteria for clinical applicability, the results in figure~\ref{fig:speed_up_factor}a suggest that there could be a time advantage in using a larger stopping criteria ($\epsilon = 10^{-3}$).

Furthermore, a single 2D Pareto surface characterization with gSA and gL-BFGS is shown in figure~\ref{fig:speed_up_factor}b. The results suggest that no significant difference in Pareto surfaces approximations is observed with GPU-based optimization engines under clinically useful stopping criteria and under more strict stopping criteria as specified in section~\ref{sec:convergence}. From these results, $\epsilon = 10^{-3}$ is used as the stopping criteria in gL-BFGS afterwards.

\subsubsection{Computational performance under clinically useful scenarios}
\begin{figure}[htbp]
\begin{subfigure}[b]{0.5\textwidth}
\centering
\includegraphics[width=\textwidth]{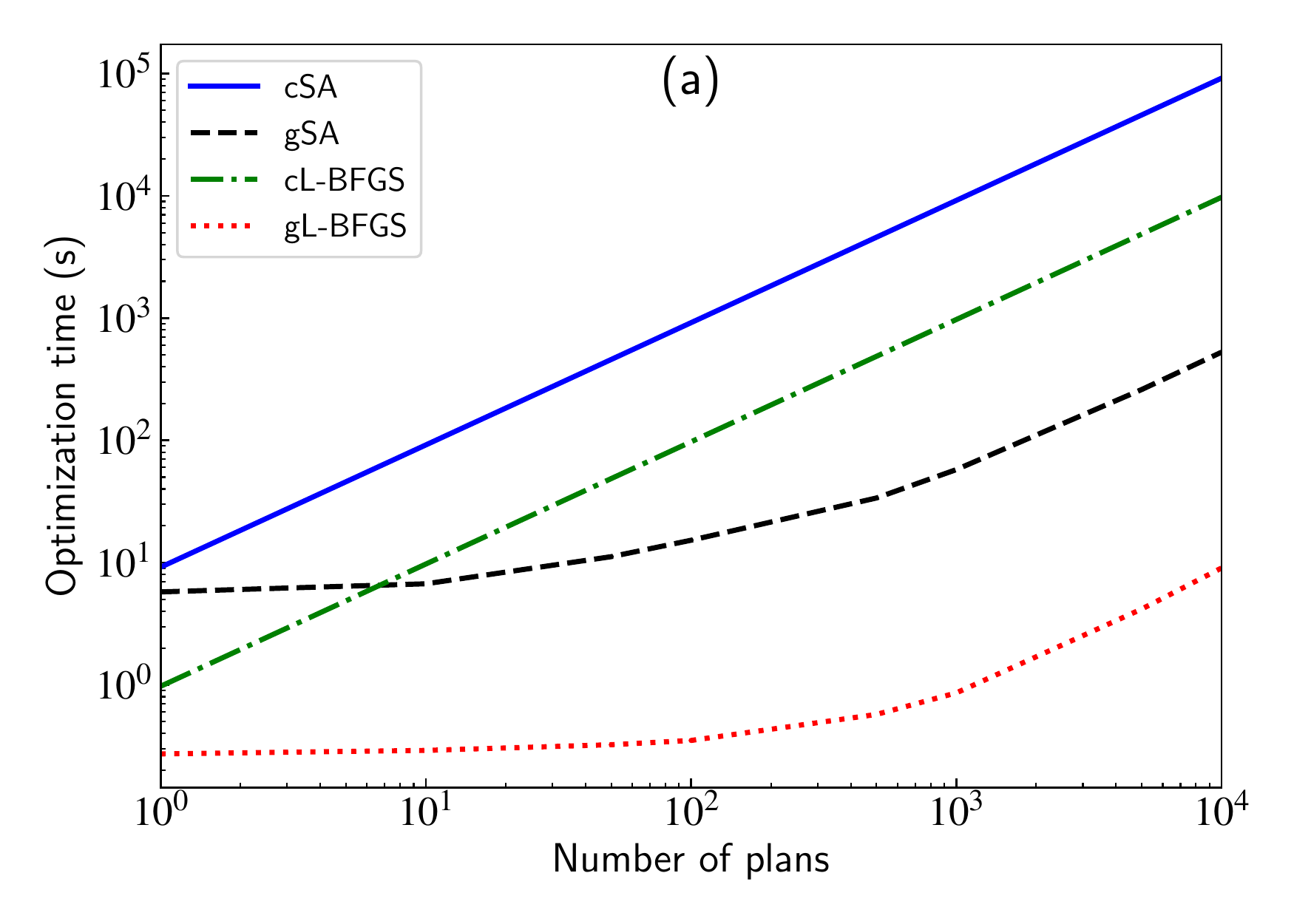}
\end{subfigure}
\begin{subfigure}[b]{0.5\textwidth}
\centering
\includegraphics[width=\textwidth]{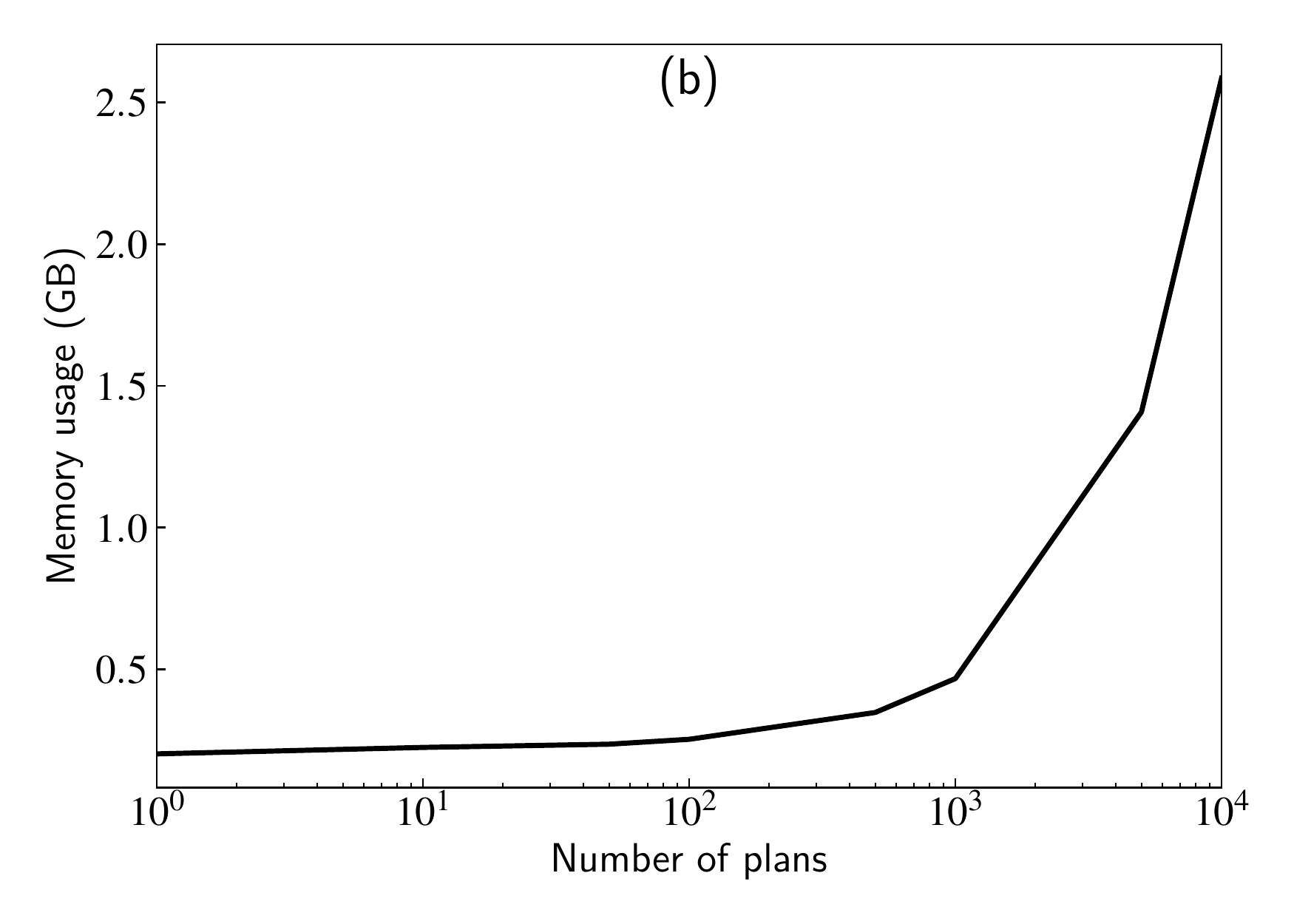}
\end{subfigure}
\caption{Computational performance against the number of plans for cSA, gSA, cL-BFGS and gL-BFGS under clinically useful scenarios (cSA and gSA: \num{1000} iterations, cL-BFGS and gL-BFGS: $\epsilon = 10^{-3}$): (a) the mean optimization time, (b) the mean GPU memory usage of gL-BFGS (the result of gSA was ignored, for its similarity to the gL-BFGS one).}
\label{fig:performance}
\end{figure}

Under clinically useful scenarios, the optimization time of cSA, gSA, cL-BFGS and gL-BFGS are shown in figure~\ref{fig:performance}a. From the results, the time of all four engines increased as the number of plans increased. For \num{1000} plans, the mean optimization time was \SI{9.2}{\second}/plan (cSA), \SI{60}{\milli\second}/plan (gSA), \SI{1}{\second}/plan (cL-BFGS), and \SI{0.9}{\milli\second}/plan (gL-BFGS). In other words, compared with the cSA result, cL-BFGS can achieve a speedup factor up to 9, gSA can achieve a speedup factor of up to 176, and gL-BFGS can achieve a speedup factor of up to \num{10990}. 

Figure~\ref{fig:performance}b shows that the mean GPU memory usage increased with the number of plans for the GPU algorithms, and that the increase rate becomes significantly large when the number of plans reaches approximately \num{1000}.

 \subsection{Patient-specific multi-criteria optimization algorithm}
\label{sec:bf_search}

\begin{figure}[htbp]
\centering
\includegraphics[width=0.6\textwidth]{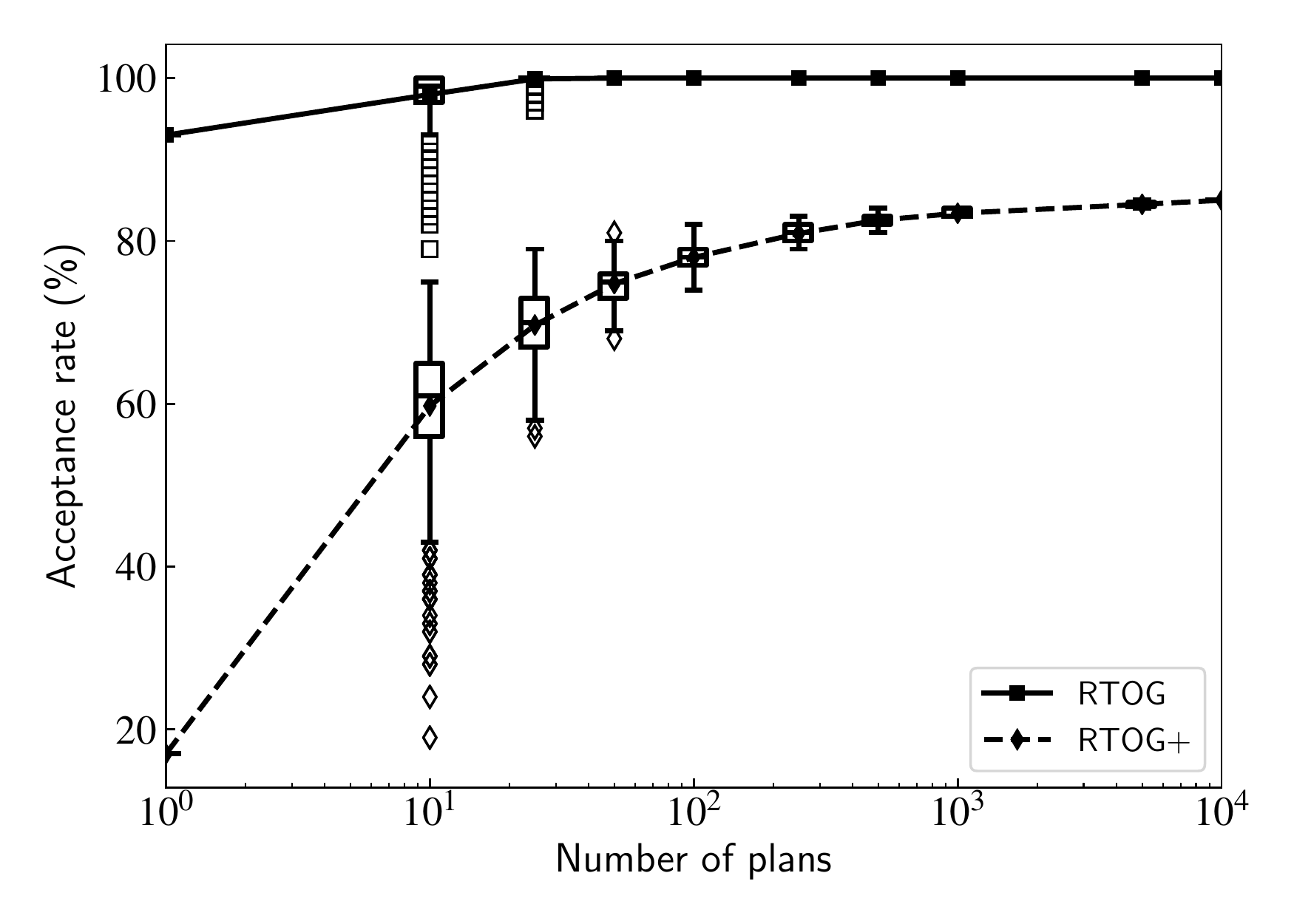}
\caption{The effect of the number of plans on RTOG and RTOG+ acceptance rates for gMCO (including the spread of the distributions in the boxes).}
\label{fig:acceptance_rate_gMCO}
\end{figure}

As the hidden weights were randomly generated in gMCO algorithm, the RTOG and RTOG+ acceptance rates were measured multiple times with different random hidden weight vectors in equation~(\ref{eq:obj}). In figure~\ref{fig:acceptance_rate_gMCO}, the RTOG+ acceptance rate increases (from 17\% to 85\%) and the spread of the acceptance rate distributions decreases with the number of plans. However, a number of \num{1000} plans was selected as the best compromised between optimization time (which increases after \num{1000} plans, see figure~\ref{fig:performance}a) and the RTOG+ acceptance rate (which does not increase significantly after \num{1000} plans) for gMCO algorithm.

\subsection{Comparison between clinical plans and gMCO plans}
\label{sec15}

\subsubsection{Dosimetric performance}

\begin{figure}[htbp]
\centering
\includegraphics[width=0.8\textwidth]{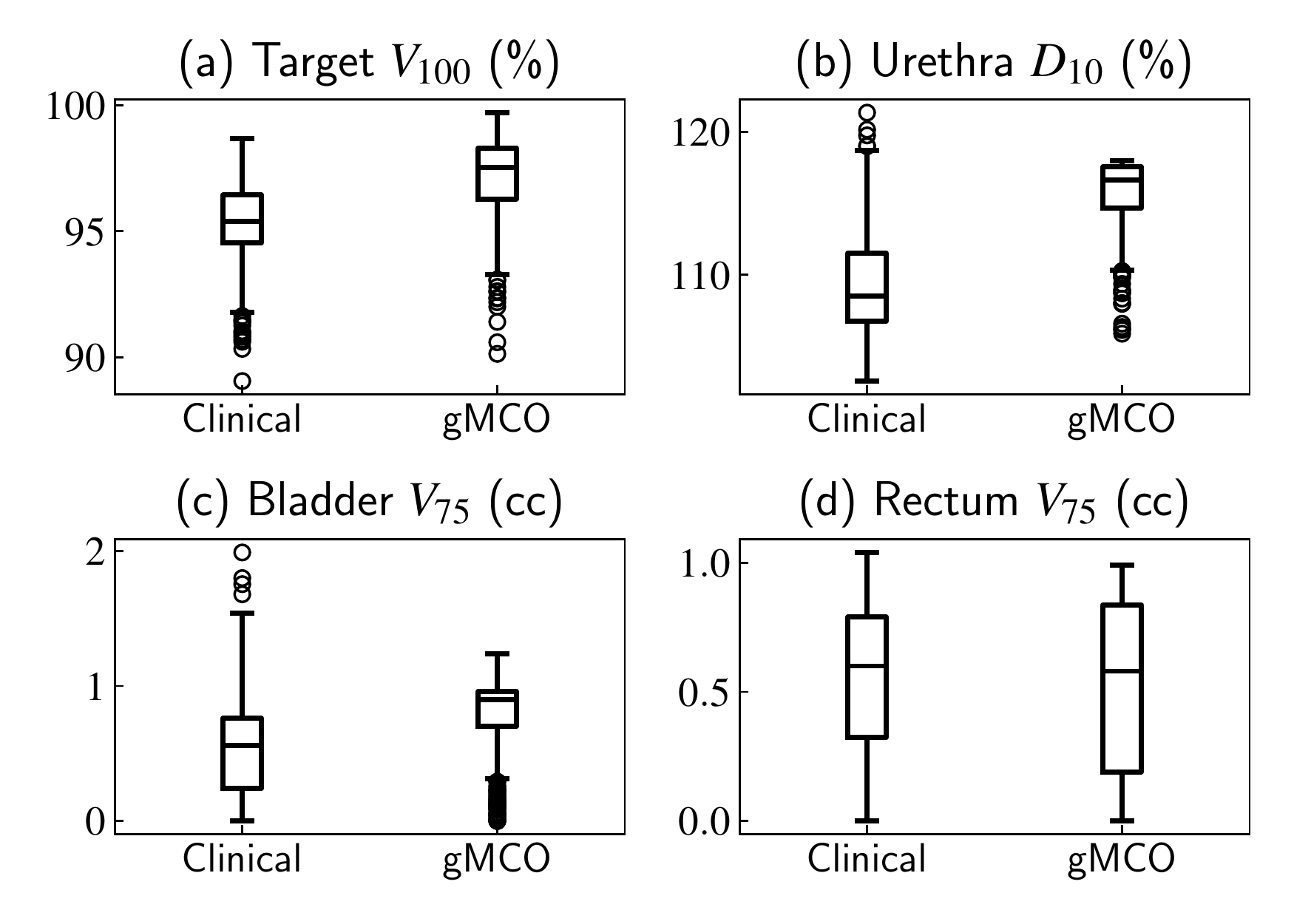}
\caption{Dosimetric comparison between IPSA physician-approved plans and gMCO plans over the test cohort: (a) target $V_{100}$, (b) urethra $D_{10}$, (c) bladder $V_{75}$, and (d) rectum $V_{75}$.}
\label{fig:comp}
\end{figure}

The dosimetric comparison between clinical plans and gMCO plans is illustrated in figure~\ref{fig:comp}. These results suggest that the mean target coverage was higher for gMCO plans (97.2\%) than for clinical plans (95.3\%). The mean urethra $D_{10}$ was significantly higher for gMCO plans (115.7\%) than for clinical plans (109.1\%). The mean bladder $V_{75}$ was 0.53~cc for clinical plans, and 0.78~cc for gMCO plans. For rectum sparing, the mean rectum $V_{75}$ was 0.56~cc for clinical plans, and 0.52~cc for gMCO plans.


The acceptance rate results are summarized in table~\ref{table:MCO}. For overall dosimetric performances, the number of RTOG valid plans was 428 (92.6\%) for clinical plans, and 461 (99.8\%) for gMCO plans. The number of RTOG+ valid plans was 288 (62.3\%) for clinical plans, and 414 (89.6\%) for gMCO plans. 

\begin{table}[htbp]
\centering
\caption{RTOG and RTOG+ acceptance rates (\%) for clinically approved plans and gMCO plans over 462 test cases.}
\label{table:MCO}
\begin{tabular}{@{}ccccccccc@{}}
\toprule
\multirow{2}{*}{} & \multicolumn{5}{c}{\textbf{RTOG}}& \multicolumn{2}{c}{\textbf{RTOG+}} & \multicolumn{1}{c}{\multirow{2}{*}{Time}} \\ \cmidrule(lr){2-6}\cmidrule(lr){7-8}
                  & \multicolumn{1}{c}{Target} & \multicolumn{1}{c}{Bladder} & \multicolumn{1}{c}{Rectum} & \multicolumn{1}{c}{Urethra} & \multicolumn{1}{c}{All} & \multicolumn{1}{c}{Target} & \multicolumn{1}{c}{All} & \multicolumn{1}{c}{} \\ \midrule
Clinical & 99.8 & 95.2 & 98.7 & 98.5 & 92.6 & 64.1 & 62.3 & mins\\ 
gMCO & 100.0 & 99.8 & 100.0 & 100.0 & 99.8 & 89.6 & 89.6 & 9.4 s \\ \bottomrule
\end{tabular}
\end{table}


The number of plans with a target coverage greater than 95\% was 296 (64.1\%) for clinical plans, and 414 (89.6\%) for gMCO plans. The number of plans that exceeded the urethra sparing constraint was 7 for clinical plans, and 0 for gMCO plans. The number of plans that exceeded the bladder sparing constraint was 22 for clinical plans, and 1 for gMCO plans. The number of plans that exceeded the rectum sparing constraint was 6 for clinical plans, and 0 for gMCO plans. In addition, the mean number of RTOG valid plans was 617/\num{1000} (61.7\%), and the mean number of RTOG+ valid plans was 268/\num{1000} (26.8\%) for the gMCO plan pool.

\begin{figure}[htbp]
\centering
\includegraphics[width=0.95\textwidth]{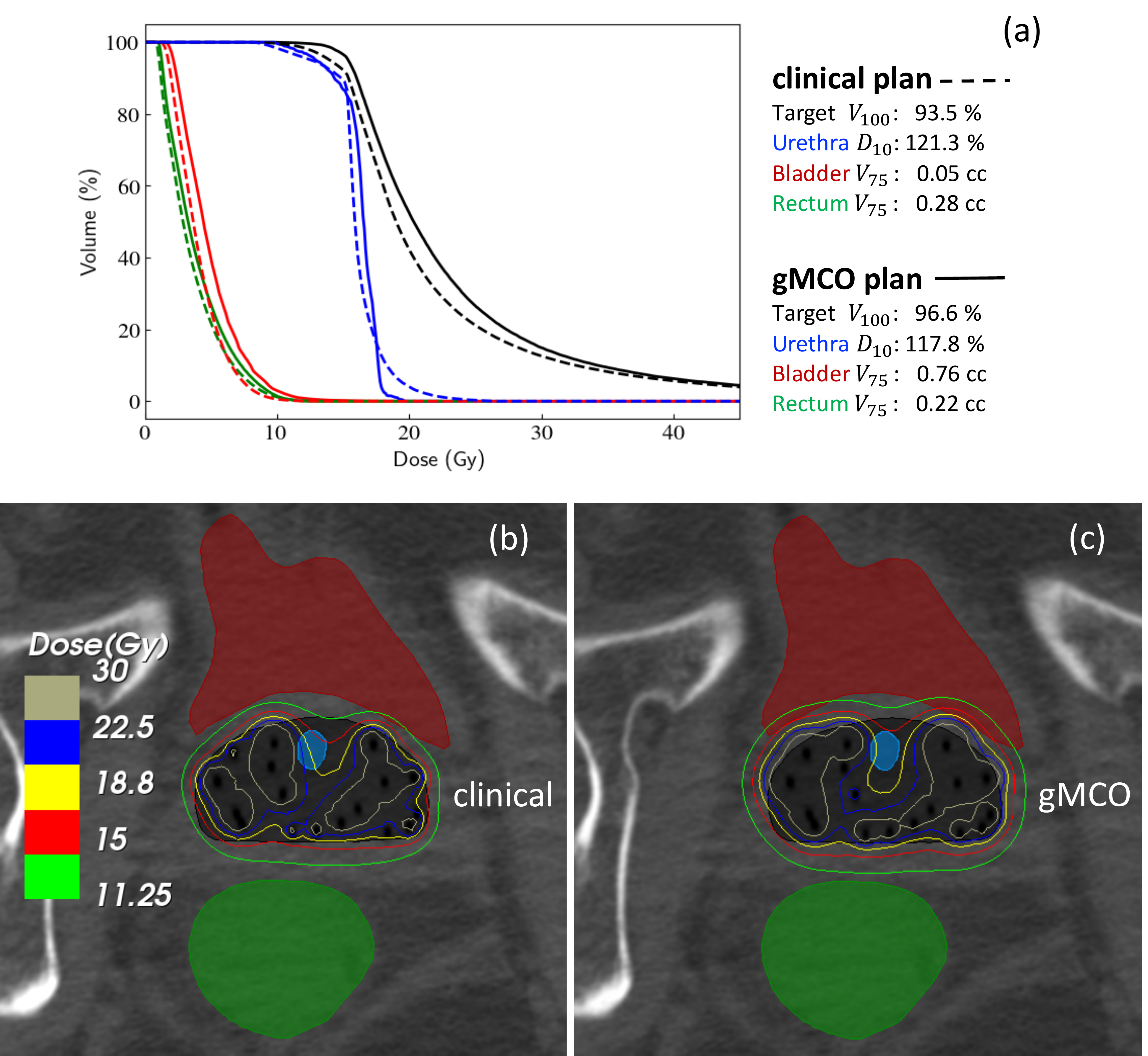}
\caption{A comparison between the clinical plan and the gMCO plan for one example case: (a) DVHs and dosimetric parameters, (b) and (c) isodose curves.}
\label{fig:example_case}
\end{figure}

As a supplement to the general comparisons described above, one example case was chosen to illustrate the advantage of gMCO in terms of the results of DVHs and isodose curves in figure~\ref{fig:example_case}.

\subsubsection{Planning time}


The time to create a plan is of the order of a few minutes in clinic, including manual tweaking of the objective function and/or dwell times. On the other hand, the mean planning time was \SI{9.4}{\second} for gMCO to generate \num{1000} optimal plans. Among these numbers, the mean dose calculation points creation time was \SI{7.4}{\second}, which represents 79\% of the mean planning time. The mean optimization time was \SI{0.8}{\second} (8.5\% of the mean total planning time). Dose rate matrix calculation and DVH calculation on GPU contribute to the rest of the mean planning time. In addition, automatically plan selection from the plan pools was performed in batch for 462 cases, and the corresponding time was negligible (\SI{4.2}{\second} for plan selection for 462 cases).

\section{Discussion}

Our recent studies~\cite{CUI18,CUI18a} showed that it is possible to obtain a RTOG valid plan without any user interventions. In order to further increase the planning efficiency, four optimization engines were implemented and compared. Our results indicated that (1) gSA and gL-BFGS can speedup the optimization time by two or three orders of magnitude compared to their CPU implementation (figure~\ref{fig:performance}a), (2) L-BFGS is equivalent to simulated annealing, and is not trapped in local minima (figure~\ref{fig:obj}), (3) gL-BFGS is able to compute \num{10000}~plans within \SI{9}{\second} (optimization time in figure~\ref{fig:performance}a), and (4) the multi-GPU approach is not necessary, considering the fact that the mean GPU memory usage to generate \num{10000} plans was 2.6 GB out of 12 GB (figure~\ref{fig:performance}b).

A new patient-specific approach called GPU-based MCO (gMCO) was proposed as an upgrade of our prior studies~\cite{CUI18,CUI18a}. gMCO can increase the RTOG acceptance rate from 97.5\%~\cite{CUI18a} to 99.8\%, and can decrease the planning time from 1~h (300 plans)~\cite{CUI18}, to \SI{41}{\second} (14 relevant plans)~\cite{CUI18a}, to \SI{9.4}{\second} (\num{1000} plans). Compared with the IPSA physician approved plans, gMCO can increase the RTOG+ acceptance rate by 27.3\%, eliminating around 10 manual tweaking needed to achieve the observed clinical level based on the results presented in figure~\ref{fig:acceptance_rate_gMCO}. For example, a RTOG invalid plan (urethra $D_{10}$ above 118\%) can be escalated to a RTOG+ valid plan by using gMCO. This has been made possible by relaxing the the bladder $V_{75}$ dose (still below 1 cc), while still meeting all requirements for target, urethra and rectum dose parameters as shown in figure~\ref{fig:example_case}a. Such information can also be seen from the isodose curves in figure~\ref{fig:example_case}b-c. Note that in this study, the trade-off involved in the automatic selection scenario is based on selecting the highest target $V_{100}$ while satisfying all the other RTOG criteria (figure~\ref{fig:comp}). However, a high quality gMCO plan pool is available for the user to pick a plan that best suits the patient-specific conditions.

KBP and MCO are widely used patient-specific inverse planning algorithms. In KBP, clinical plans were used to extract the regression models based on geometric features. However, clinical plans are user-dependent~\cite{DAS08}, and may be inconsistent between centers~\cite{CHUN08}. On the other hand, gMCO is independent of these issues. In MCO, even though interpolations between calculated plans were usually used to achieve a high planning efficiency, ultra-fast planning remains a challenge since no parallelization scheme was implemented. In this study, it only takes \SI{9.4}{\second} to generate a high quality plan pool with gMCO. However, it is admitted that these comparisons were made by ignoring that the dwell times optimization in HDR brachytherapy is a relatively small scale problem compared to the fluence map optimization in EBRT.

Note that for the objective function considered in this work, the solution space is convex and it would be easy to dismiss SA in favor of the more computationally efficient gL-BFGS algorithm. While this objective function is popular in the field, other types of objective function might have more complex solution spaces. Therefore, having a robust, albeit slower, MCO algorithm based on SA remains an essential tool.

We anticipate that the approach proposed in this study will be implemented in clinical systems as an adjunct tool. In future work, the application of gL-BFGS as well as gMCO to other HDR brachytherapy sites will be investigated.

\section{Conclusion}

Two GPU-based optimization engines were designed to calculate multiple plans in parallel. With the preferred engine, an ultra-fast patient-specific planning tool that is able to generate a high quality plan without any user interventions was proposed. After a validation over a large-scale patient cohort, both plan quality and planning efficiency can be significantly improved compared with the traditional planning in clinic.

\section*{Acknowledgement}

This work was supported in part by the National Sciences and Engineering Research Council of Canada (NSERC) via the NSERC-Elekta Industrial Research Chair Grant (\#484144-15), via the NSERC Discovery Grants (\#355493 and \#435510), and via the CREATE Medical Physics Research Training Network Grant (\#432290). The authors acknowledge a scholarship from the Chinese Scholarship Council and partial support by the Canada Foundation for Innovation (\#CFI30889). The authors acknowledge the supports from their colleagues, especially Louis Archambault, Paul Edimo, Andrea Frezza, Charles Joachim-Paquet, Fr\'ed\'eric Lacroix, Marie-Claude Lavall\'ee, Ghyslain Leclerc, Lo\"ic Paradis-Laperri\`ere, \'Eric Poulin, and Nicolas Varfalvy.

\section*{References}
\bibliography{iopart-num}
\end{document}